# Stable magnesium peroxide at high pressure

Sergey S. Lobanov[1,2,*], Qiang Zhu[3], Nicholas Holtgrewe[1,4], Clemens Prescher[5], Vitali B. Prakapenka[5], Artem R. Oganov[3,6,7], Alexander F. Goncharov[1,8,9]

[1]Geophysical Laboratory, Carnegie Institution of Washington, Washington, DC 20015, USA

[2]V.S. Sobolev Institute of Geology and Mineralogy SB RAS, Novosibirsk 630090, Russia

[3]Department of Geosciences, Department of Physics and Astronomy, Stony Brook University, Stony Brook, NY 11794, USA

[4]Howard University, 2400 Sixth Street NW, Washington, DC 20059, USA

[5]Center for Advanced Radiation Sources, University of Chicago, Chicago, IL 60632, USA

[6]Moscow Institute of Physics and Technology, 9 Institutskiy lane, Dolgoprudny city, Moscow Region, 141700, Russian Federation

[7]School of Materials Science, Northwestern Polytechnical University, Xi'an, 710072, China

[8]Key Laboratory of Materials Physics, Institute of Solid State Physics, CAS, Hefei, 230031, China

[9]University of Science and Technology of China, Hefei, 230026, China

*slobanov@carnegiescience.edu

**Abstract**

Rocky planets are thought to comprise compounds of Mg and O as these are among the most abundant elements, but knowledge of their stable phases may be incomplete. MgO is known to be remarkably stable to very high pressure and chemically inert under reduced condition of the Earth's lower mantle. However, in 'icy' gas giants as well as in exoplanets oxygen may be a more abundant constituent[1, 2]. Here, using synchrotron x-ray diffraction in laser-heated diamond anvil cells, we show that MgO and oxygen react at pressures above 94 GPa and T = 2150 K with the formation of the theoretically predicted *I4/mcm* $MgO_2$ (Ref.[3]). Raman spectroscopy detects the presence of a peroxide ion ($O_2^{2-}$) in the synthesized material as well as in the recovered specimen. Likewise, energy-dispersive x-ray spectroscopy confirms that the recovered sample has higher oxygen content than pure MgO. Our finding suggests that $MgO_2$ may substitute MgO in rocky mantles and rocky planetary cores under highly oxidizing conditions.

**Main Text**

Oxygen and magnesium are the first and second most abundant elements in the Earth's mantle[4]; thus knowledge of stable phase relations in the Mg-O system as a function of thermodynamic parameters is necessary input information for reconstructing Earth-like planetary interiors. For example, ferropericlase (MgO with a relatively low Fe content) is the second most abundant mineral on Earth owing to its remarkable thermodynamic stability in the *Fm3m* crystal structure (up to 500 GPa and at least 5000 K for pure MgO)[5, 6]. This is why ferropericlase has been assumed in gas giant cores[7, 8] as well as in extrasolar terrestrial mantles[9, 10]. However, planet-harboring stars vary in chemical composition[11], which likely affects the composition of planetary building blocks and exoplanet mineralogy[2]. Therefore, Earth-like mantle mineralogy should not be assumed for terrestrial exoplanets. Elevated oxygen contents have been observed in planet-host stars[1], which may affect the stability of MgO and favor other solid phases in the Mg-O system[3, 12]. For example, magnesium peroxide ($MgO_2$) may be synthesized at near-ambient conditions and at high oxygen fugacities in the pyrite-type (*Pa3*) structure[12]. However, *Pa3* $MgO_2$ is thermodynamically unstable and readily decomposes to MgO and $O_2$ upon heating to 650 K at ambient pressure[12]. The intrinsic instability of $MgO_2$ is attributed to the strong polarizing effect of the $Mg^{2+}$ ion possessing high charge density in a relatively small ionic radius[13]. This is why the stability of Group II peroxides increases down the Group: beryllium peroxides are not known[13], while Ca, Sr and Ba form increasingly more stable peroxides at ambient conditions[14, 15]. Therefore, using empirical considerations on chemical pressure[16, 17] $MgO_2$ may be expected to become stable under high pressure conditions. Indeed, *ab initio* simulations found that *I4/mcm* $MgO_2$ becomes stable at P > 116 GPa (Ref.[3]) and 0 K. Here, we report on the synthesis of *I4/mcm* $MgO_2$ in laser-heated diamond anvil cell (DAC). $MgO_2$ may be an abundant mineral in the interiors of Uranus, Neptune, and highly oxidized terrestrial exoplanets. Our finding also suggests that the Mg-Fe-Si-O system likely has more unexpected chemistry at high pressure.

Two types of chemical precursors were loaded in DAC to study the $MgO-O_2$ phase diagram in the 0-154 GPa pressure range (see Table 1 and Methods). In type-A experiments we put two 4 μm thick MgO disks in the sample cavity which was subsequently filled with liquefied oxygen (Fig. 1, inset). In type-B runs we used commercially available magnesium peroxide complex (24-28% *Pa3* $MgO_2$, 42-46% MgO, ~30% Mg) mixed with submicron Au powder serving as a laser absorber. The mixture was loaded without pressure medium.

*Table 1. Experiments description.*

| Type | # | Precursor | Culet size, μm | Maximum pressure, GPa | *I4/mcm* $MgO_2$ | Pressure calibrant |
|---|---|---|---|---|---|---|
| A | 1 | $MgO+O_2$ | 300/80 | 95 | Yes | MgO |
| A | 2 | $MgO+O_2$ | 200 | 103 | Yes | MgO |
| B | 1 | $MgO+O_2+MgO_2$ | 200 | 70 | No | MgO, Au |
| B | 2 | $MgO+O_2+MgO_2$ | 300/100 | 154 | Yes | MgO, Au |

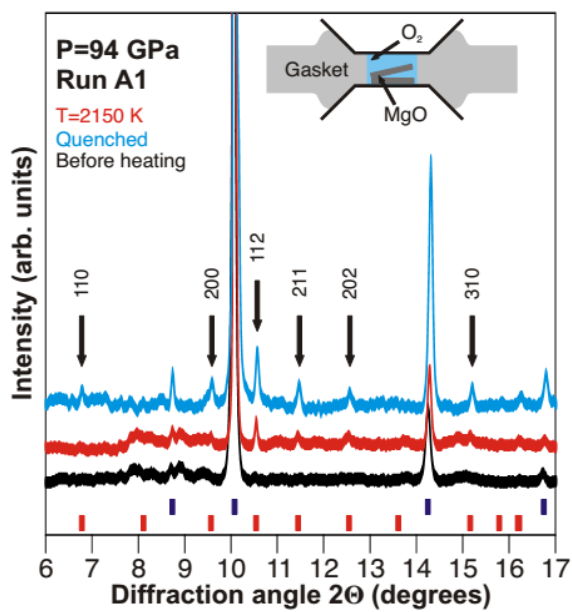

*Figure 1.* X-ray diffraction (XRD) pattern of the A1 sample before laser heating (black line), at high temperature (red line) and quenched to 300 K (blue line). Arrows mark new peaks that appear at high temperature. A thermal shift of the MgO peaks is seen at T = 2150 K indicating a uniform heating of the sample. Miller indices correspond to the indexed tetragonal unit cell. Positions of the Bragg reflections of *I4/mcm* $MgO_2$ (Ref.[3]) are shown by red ticks. Blue bars correspond to MgO. Oxygen peaks are not resolved. The wavelength is 0.3344 Å. The inset shows the experimental assemblage of type-A runs.

**Synchrotron x-ray diffraction (XRD)**

Figure 1 shows representative XRD patterns of the run A1 at 94 GPa before heating, at 2150 K, and after quenching. Oxygen peaks are weak and not resolved in the integrated pattern, although some of them are visible in the caked XRD images forming a spotty pattern (Supplementary Fig. S1). Six new peaks appear upon heating and become clearly seen in the XRD pattern of the quenched sample. Indexing the new peaks reveals a tetragonal unit cell with $a = 4.000(1)$ Å, $c = 4.743(5)$ Å. The new peaks show a good match with the expected positions of the predicted *I4/mcm* $MgO_2$ Bragg reflections[3] (shown as red ticks in the Fig. 1). Rietveld refinement of the new phase was not possible due to its spotty diffraction texture and because low intensity peaks could not be resolved.

In the experiments with type-B precursors MgO, $\varepsilon$-$O_2$, and Au were the only phases observed in XRD patterns after it was heated to T > 2000 K in the pressure range of 5-70 GPa. Bragg peaks that can be assigned to *Pa3* $MgO_2$ were completely absent in the reaction products suggesting that the precursor had decomposed to MgO and $O_2$. Indeed, the presence of pure oxygen in the quenched sample was confirmed with Raman spectroscopy. Noteworthy, we did not observe elemental Mg (neither *hcp* at P < 50 GPa nor *bcc* at P > 50 GPa) in the reaction products. Magnesium likely reacts with oxygen as the latter gets liberated upon *Pa3* $MgO_2$ decomposition at high temperature.

Laser heating of the B2 sample to T > 2000 K at P = 130 GPa provided more information on the high pressure chemistry of the Mg-O system. We were very curious to note that new peaks form a powder-type texture in XRD images (Supplementary Fig. S2), indicating the presence of a large number of randomly oriented crystallites. Surprisingly, the spotty texture is now built by MgO and $\zeta$-$O_2$. Indexing the most clearly resolved new peaks again yields a tetragonal unit cell with $a = 3.925(1)$ Å, $c = 4.613(6)$ Å. Moreover, the obtained Miller indices reproduce that of the tetragonal phase synthesized in the A1 run (Fig. 1) suggesting that the exact same phase has been produced in the A1 and B2 runs. Given the large yields of the new phase as well as the polycrystalline sample texture, Rietveld method can be applied to test and refine the theoretically predicted *I4/mcm* $MgO_2$. According to the prediction by Zhu et al. (Ref.[3]), magnesium occupies a 4a Wyckoff position (*0, 0, 0.25*) and oxygen is located in 8h (*x, x+0.5, 0*), which leaves only the *x* fractional coordinate of oxygen to refine. The refined $x = 0.1285(13)$, and the predicted $x = 0.126$ is within $2\sigma$; thus the refined structural model may be considered identical to the predicted one. Figure 2 compares the experimental XRD pattern with the synthetic XRD of the Rietveld-refined *I4/mcm* $MgO_2$. Refined lattice parameters (Supplementary Table S1) were used to construct the *I4/mcm* $MgO_2$ P-V equation of state (EOS) (Fig. 3). We also computed the $MgO_2$ volume in the 70-150 GPa pressure range (Supplementary Table S2). The EOS parameters are reported in the Supplementary Table S3.

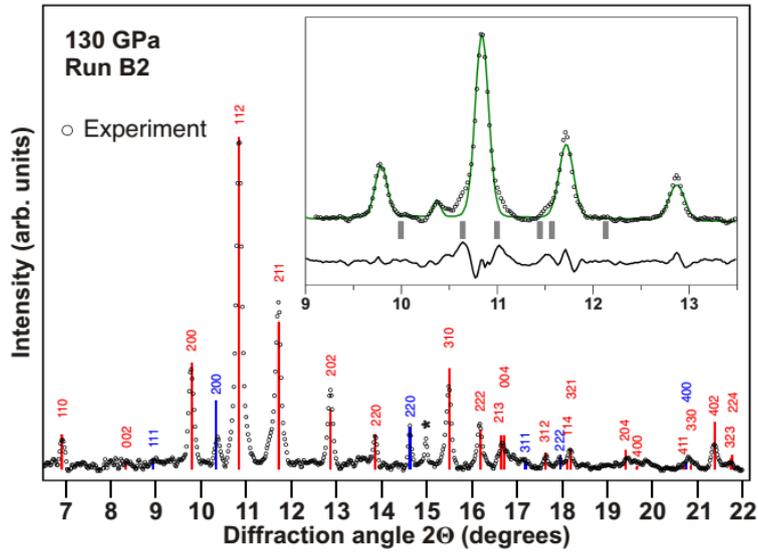

*Figure 2.* XRD of the type-B precursor laser-heated to T > 2000 K at 130 GPa. Red bars represent positions and intensities of Bragg reflections of the Rietveld refined I4/mcm $MgO_2$. Dark blues bars correspond to MgO. The peak marked with an asterisk belongs to oxygen. **Inset:** Rietveld refinement of the $MgO_2$ crystal structure. Grey bars approximate positions of the strongest $\zeta$-$O_2$ peaks[18]. Green curve represents the calculated intensities of the refined structure ($I_{calc}$). Black line is the intensity difference curve ($I_{obs}$-$I_{calc}$). Calculated residuals after background subtraction are $R_{exp}$=0.138, $R_{wp}$=0.265; calculated following Ref.[19]. The x-ray wavelength is 0.3344 Å.

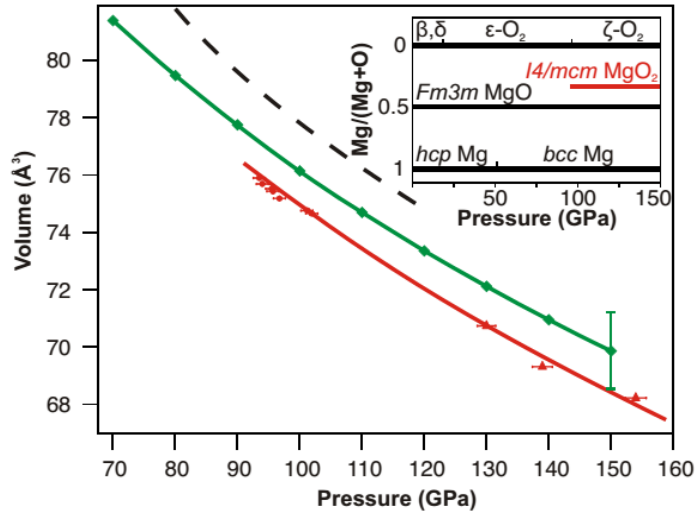

*Figure 3.* The 300 K third-order Birch-Murnaghan EOS of I4/mcm $MgO_2$. Red line is EOS fit to the experimental data from runs A1, A2 (red circles), and B2 (red triangles).The pressure error bar is based on the reported uncertainty of the MgO EOS[20]. Green diamonds and green line are the DFT EOS of I4/mcm phase. Black dashed line is the sum of the unit cell volumes of MgO and $O_2$ (taken with proper coefficients as dictated by the *synthesis reaction and the number of formula units in the MgO and O₂ unit cells). Inset: Experimental pressure-composition phase diagram of the Mg-O system as determined in this work.*

**Raman spectroscopy** was applied to characterize *I4/mcm* $MgO_2$, albeit the increased fluorescent background of diamond anvils typical at pressures exceeding 100 GPa. On top of this, oxygen becomes metallic at pressure above 96 GPa[18] and may screen reaction products from the probe laser radiation. First, we used density-functional perturbation theory (DFT) to compute spectral position and intensities of *I4/mcm* $MgO_2$ Raman bands in the 90-150 GPa pressure range. Group theory for the *I4/mcm* $MgO_2$ allows 5 Raman active vibrations ($2E_g + B_{1g} + A_{1g} + B_{2g}$). Our DFT calculations suggest that $B_{2g}$ and $A_{1g}$ modes should have observable intensities with $A_{1g}$ being the most intense as it may also be anticipated from earlier Raman studies of solid peroxides[21]. Figure 4A shows Raman spectra of A2 at 103 GPa collected from an area containing *I4/mcm* $MgO_2$ as established by XRD. The $O_2$ vibron was also observed in Raman spectra collected from the laser-heated spot. Since both ε- and ζ-$O_2$ have a rich Raman spectrum[22] at frequencies lower than 900 cm⁻¹ it is difficult to use this spectral region for a reliable identification of the *I4/mcm* $MgO_2$. Luckily, the position of $A_{1g}$ band is predicted in the 1060-1175 cm⁻¹ spectral range at 90-150 GPa according to our DFT calculations. Based on this comparison, the high-frequency mode at 1037 cm⁻¹ may be assigned to the O-O stretching vibration in the peroxide ion. Raman shift of the high-frequency band is in agreement with the positions of $A_{1g}$ band in $H_2O_2$ (Ref.[23]) and $BaO_2$ (Ref.[24]) confirming the assignment.

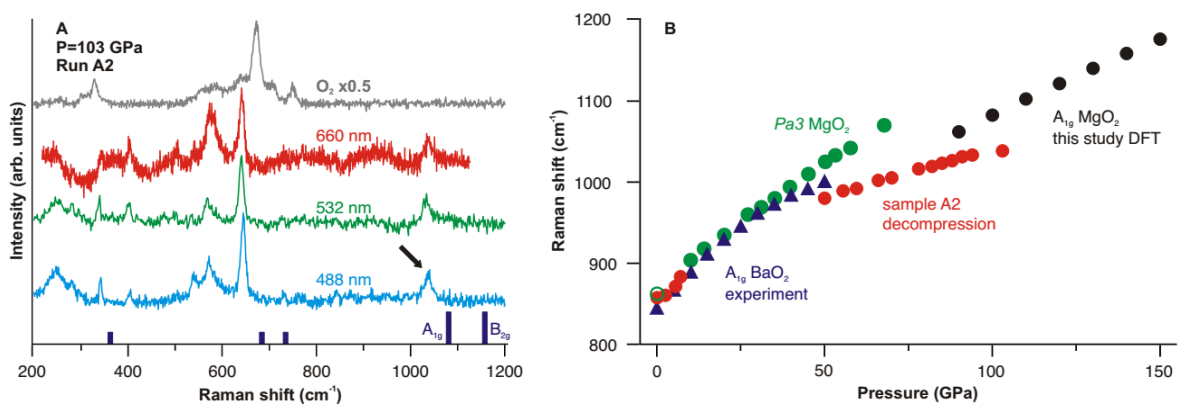

***Figure 4.*** *(**A**) Raman spectra of MgO + $O_2$ reaction products collected with 488, 532, and 660 nm excitations. Oxygen Raman spectra collected outside of the laser-heated region is shown for comparison. Dark blue vertical ticks correspond to the computed Raman modes of I4/mcm at 100 GPa ($A_{1g}$ and $B_{2g}$ modes may have observable intensities, according to our DFT computations). The 1037 $cm^{-1}$ peak that can be assigned to the $A_{1g}$ mode in I4/mcm $MgO_2$ is marked by an arrow. (**B**) Pressure dependencies of O-O symmetric stretching vibration ($A_{1g}$). Red circles represent positions of the high-frequency mode observed in the A2 sample. Green circles correspond to the positions of the $A_{1g}$ band in Pa3 $MgO_2$ measured in this study and in Ref.[21] (open circle at 1 atm.). Blue triangles are positions of the $A_{1g}$ mode in $BaO_2$. Black circles are computed frequencies of the $A_{1g}$ mode in I4/mcm $MgO_2$.*

Raman spectra of *I4/mcm* $MgO_2$ were followed on A2 decompression run. In Figure 4B the pressure-frequency dependence of the $A_{1g}$ band of *I4/mcm* $MgO_2$ is compared with that in $BaO_2$ (Ref.[24]) and *Pa3* $MgO_2$ (this study and Ref.[21]). We could only trace the high-frequency band down to 50 GPa, and then at 0-10 GPa because of the overlap with the overtone of oxygen L2 peak $(2\upsilon_{L2})$[22]. Expectedly, the frequency-pressure dependence of the O-O symmetric stretching in *Pa3* $MgO_2$ is similar to that in *I4/mmm* $BaO_2$. The DFT-computed frequencies of the $A_{1g}$ in *I4/mcm* $MgO_2$ also form a similar slope in the 90-150 GPa pressure range. However, the measured pressure dependence of the high-frequency band in the synthesized sample is less steep. Interestingly, at 1 bar the position of high-frequency band (857 $cm^{-1}$) is almost identical to the position of $A_{1g}$ mode in *Pa3* $MgO_2$ (864 $cm^{-1}$)[21] suggesting that the recovered product is likely *Pa3* $MgO_2$. Overall, our data provide spectroscopic evidence for the peroxide ion in the synthesized material and that the material containing peroxide ion is preserved to ambient conditions.

**Stability of *I4/mcm* $MgO_2$ and planetary implications**

*I4/mcm* $MgO_2$ can be synthesized in the mixture of MgO with $O_2$ at 94GPa indicating a thermodynamic stability of $MgO_2$ at this pressure, which is close to the theoretical predicted pressure of 116 GPa (Ref.[3]). We therefore conclude that *I4/mcm* $MgO_2$ is a thermodynamically stable phase in the high pressure phase diagram of the Mg-O system (Fig. 3, inset). On decompression, we could only follow the new phase in XRD down to 74 GPa, while Raman spectroscopy shows no discontinuities in the position of the high-frequency band down to at least 50 GPa. It remains unclear what physicochemical transformations occurred in the synthesized phase at P < 50 GPa, but at 1 bar the laser-heated area of the recovered sample (A2) (Supplementary Fig. S3) shows a Raman signature of $P_{a3}$ $MgO_2$. Mapping the extracted sample with an energy-dispersive x-ray spectroscopy (EDS) revealed that the laser-heated area has higher oxygen content than the area that was not subjected to high temperatures (Supplementary Fig. S4). Detailed chemical characterization, however, was not possible because unreacted MgO is mixed with the oxygen-rich phase in the laser-heated area. Nevertheless, EDS analysis provides independent evidence for $MgO_2$ in the recovered sample.

The thermodynamic stability of *I4/mcm* $MgO_2$ at P > 94 GPa is not surprising as heavier Group II elements, strontium and barium, form stable peroxides with $CaC_2$-type (*I4/mmm*) crystal structure at ambient pressure with the O-O bond parallel to the *c* axis and 2 $MO_2$ (M=Sr, Ba) formula units in the unit cell[14]. The O-O chemical bond length in $MgO_2$ is 1.454(1) Å at 93.5 GPa, which is comparable to that of $SrO_2$ (1.483 Å) and $BaO_2$ (1.493 Å) at standard conditions[14]. *I4/mcm* $MgO_2$, however, has 4 formula units in the unit cell and the O-O bond is parallel to the *ab* plane diagonal (Supplementary Fig. S5 A, B).

Taking into account that *Fm3m* MgO has 4 formula units and *C2/m* oxygen (ε-, ζ-) has 8 $O_2$ molecules in the unit cell we calculated the volume of MgO + 1/2 $O_2$ as a function of pressure using the reported MgO and $O_2$ EOS[18, 20] (Fig. 3, dashed line). It is apparent that *I4/mcm* $MgO_2$ is denser than the reactants in the studied pressure range. Interestingly, the reaction of MgO with $O_2$ at P > 94 GPa promotes an 8-fold coordination of $Mg^{2+}$ at much lower pressures than expected for *Fm3m* to *Pm3m* (NaCl-type to CsCl-type) transition in pure MgO (~500 GPa)[3, 5, 6]. Because of the covalent O-O bond one may expect an increased electron density in between the oxygen atoms. In this scenario, $Mg^{2+}$ may form an ionic bond with the $O_2^{2-}$ dumbbell centers, in which case the *I4/mcm* MgO$_2$ structure may be viewed as the CsCl-type structure (Supplementary Fig. S5 B, C).

*In situ* XRD at T = 2150 K (Fig. 1) demonstrates that MgO$_2$ is stable at high temperature. Thus, MgO$_2$ may substitute MgO in planetary interiors in the presence of free oxygen in terrestrial exoplanets with elevated oxygen content (compared to Earth)[1, 2]. Likewise, dissociated water[25-27], the most abundant planetary ice, may be expected to react with rocky material present in the mantles and cores of the icy giant planets such as Uranus and Neptune with the formation of *I4/mcm* MgO$_2$. Such chemical interaction may result in chemical and physical stratification in the interiors of Uranus and Neptune, which is believed to be responsible for the peculiar non-axisymmetric non-dipolar magnetic fields[28]. Overall, the case of *I4/mcm* MgO$_2$ shows that chemical transformation of minerals at high pressure may play an important and yet unrevealed role in planetary forming systems such as Mg-Fe-Si-O.

**Methods**

**Materials and samples**

Diamond anvils with culets of 200, 300/100, and 300/80 μm were used to access the 100-150 GPa pressure range. Rhenium foils (200 μm thick) were indented to a thickness of 30-40 μm and then laser-drilled to create holes (30-100 μm in diameter) serving as sample chambers. Magnesium oxide (99.85%) available from Alfa-Aesar was used for the type-A experiments. Before sample loadings magnesia was annealed at 1293 K for 12 hours to get rid of any adsorbed water. Two MgO disks were made by compressing the magnesia powder to a thickness of 4-5 μm and were stacked in the gasket hole. The remaining volume of the sample chamber was filled with liquefied zero-grade oxygen (99.8%, Matheson Gas Products) at approximately 77 K. In type-B experiments we used magnesium peroxide complex available from Sigma-Aldrich (24-28% *Pa3* MgO$_2$, 42-46% MgO, ~30% Mg). The magnesium peroxide complex was mixed with submicron gold powder and loaded in the sample chambers with no pressure medium.

**Synthesis and characterization**

All XRD experiments were performed at the undulator beamline at 13ID-D GeoSoilEnviroCARS, APS, using the online double-sided laser-heating system[29]. The laser-heating radiation was coupled to oxygen in type-A, and to the gold powder in type-B runs. Synchrotron XRD was collected in-situ at high temperature and high pressure in the diamond anvil cells to determine the onset of chemical and physical transformations with the x-ray beam (37.077 keV) focused to 4 μm spot size. Quenched samples were mapped with the x-ray beam with a step size of 5 (A runs) or 2 (B runs) μm to find areas with less Au and $O_2$, which was necessary for a better $MgO_2$ characterization. Temperature was measured simultaneously with XRD spectroradiometrically and calculated using the T-Rax software (C. Prescher). Pressure was determined using the MgO equation of state[20] with an uncertainty of less than 1%. In type-B experiments pressure was double-checked using the Au EOS[30].

Raman characterization of the quenched samples was performed in the Geophysical Laboratory. Solid-state lasers with 488, 532, and 660 nm lines were used as excitation sources. Backscattered Raman radiation was analyzed by a single-stage grating spectrograph equipped with a CCD array detector. The spectral resolution was 4 $cm^{-1}$.

Energy-dispersive x-ray spectroscopy (EDS) analysis was performed on a dual beam focused ion beam / scanning electron microscope (FIB/SEM Zeiss Auriga 40) equipped with an Oxford X-Max 80 $mm^2$ large-area silicon drift detector at the accelerating voltage of 5kV in the Geophysical Laboratory. The analyzed sample was coated with Ir (~ 5 nm) to prevent specimen charging. Pyrope and the ENEL20 glass were used as standards for oxygen and magnesium, respectively.

Details of the XRD analysis and Computational Methods as well as appropriate references are provided in the Supplementary Information.

**Acknowledgments:** The study was supported by the Deep Carbon Observatory, the National Science Foundation (EAR-1114313, EAR-1015239, EAR-1128867, DMR-1231586), DARPA (Grants No. W31P4Q1210008 and No. W31P4Q1310005), the Government of Russian Federation (grants 14.A12.31.0003 and 14.B25.31.0032), and Foreign Talents Introduction and Academic Exchange Program (No. B08040). Calculations were performed on XSEDE facilities and on the cluster of the Center for Functional Nanomaterials, Brookhaven National Laboratory, which is supported by the DOE-BES under contract no. DE-AC02-98CH10086. Maddury Somayazulu and other co-workers at the Geophysical Laboratory are thanked for their comments on earlier versions of this manuscript.

**Author Contributions:** S.S.L., A.F.G., and A.R.O. designed the study. S.S.L., N.H., A.F.G. performed the experimental work with an active assistance of C.P. and V.B.P. Theoretical calculations were performed by Q.Z., and A.R.O. All authors discussed the results and the implications. S.S.L. analyzed the data and wrote the paper.

# SUPPLEMENTARY MATERIAL

# Stable magnesium peroxide at high pressure

Sergey S. Lobanov[1,2,*], Qiang Zhu[3], Nicholas Holtgrewe[1,4], Clemens Prescher[5], Vitali B. Prakapenka[5], Artem Oganov[3,6,7], Alexander F. Goncharov[1,8,9]

[1]Geophysical Laboratory, Carnegie Institution of Washington, Washington, DC 20015, USA

[2]V.S. Sobolev Institute of Geology and Mineralogy SB RAS, Novosibirsk 630090, Russia

[3]Department of Geosciences, Department of Physics and Astronomy, Stony Brook University, Stony Brook, NY 11794, USA

[4]Howard University, 2400 Sixth Street NW, Washington, DC 20059, USA

[5]Center for Advanced Radiation Sources, University of Chicago, Chicago, IL 60632, USA

[6]Moscow Institute of Physics and Technology, 9 Institutskiy lane, Dolgoprudny city, Moscow Region, 141700, Russian Federation

[7]School of Materials Science, Northwestern Polytechnical University, Xi'an, 710072, China

[8]Key Laboratory of Materials Physics, Institute of Solid State Physics, CAS, Hefei, 230031, China

[9]University of Science and Technology of China, Hefei, 230026, China

*slobanov@carnegiescience.edu

### *I4/mcm* MgO$_2$ equation of state

Bragg peaks of MgO$_2$ were sharp in quenched samples right after the synthesis which allowed for a reliable volume determination with small σ values (e.g. run A2). On decompression, however, XRD peaks become broad and volume measurements were less certain (B2). The two P-V datasets were fit with a room temperature third-order Birch-Murnaghan equation of state (EOS)[1]:

$$P = \frac{3}{2}K_0 \left[\left(\frac{V_0}{V}\right)^{\frac{7}{3}}\right] - \left[\left(\frac{V_0}{V}\right)^{\frac{5}{3}}\right]\left\{1 - \frac{3}{4}(4 - K_0`)[\left(\frac{V_0}{V}\right)^{\frac{2}{3}} - 1]\right\},$$

where $K_0$ and $K_0`$ are the bulk modulus and its pressure derivative, respectively; $V_0$ and $V$ are the unit cell volumes at standard conditions and at high pressure, respectively. The new phase was still observed in XRD of the sample B2 decompressed down to 74 GPa, although the unit cell measurements were not reliable as peaks broaden probably due to the phase instability at this pressure. P-V data obtained on the sample B2 decompression is marked with an asterisk in the Supplementary Table S1. At P < 74 GPa the XRD peaks become too broad and start overlapping with peaks from other materials precluding identification of the MgO$_2$ phase. The theoretically computed volumes are systematically 1.6-2.1 % larger than the experimental ones in the 100-150 GPa pressure range, which is within the computational uncertainty.

### X-ray diffraction analysis

2D XRD patterns were integrated using the Dioptas software (written by C. Prescher). Manual background subtraction was done in Fityk[2]. Preliminary Bragg peaks indexing was performed with Dicvol06 (Ref.[3]). GSAS/EXPGUI[4, 5] was used for Rietveld refinement in accordance with the guidelines provided in Ref.[6]. Oxygen spotty reflections overlapping with the continuous lines produced by the new phase were masked. Also, we did not use the region of 2θ > 13° where the background scattering is not uniformly distributed in the azimuth range of 0 to 360°. Scaling factors and unit cell parameters were refined first. Subsequently, peak profiles were fit with the pseudo-Voigt function and, at last, we refined the oxygen fractional coordinate ($x$ in the 8h position). Crystal structures were visualized with the use of VESTA 3 package[7]. The 300 K third-order Birch-Murnaghan equation of state (EOS) was obtained using a (sigma)volume-weighted fitting procedure was performed as implemented in the EoSFit7GUI (R. Angel).

**Computational Methods**

Density functional theory (DFT) within the Perdew-Burke-Ernzerhof (PBE) generalized gradient approximation (GGA)[8] as implemented in the VASP code[9], was used for structural and vibrational analysis. For the structural relaxation, we used the all-electron projector-augmented wave (PAW) method[10] and the plane wave basis set with the 600 eV kinetic energy cutoff; the Brillouin zone was sampled by Γ-centered meshes with the resolution $2\pi \times 0.06$ A$^{-1}$. The phonon frequencies were calculated using the finite displacement approach as implemented in the Phonopy code[11]. The Raman intensities were obtained by computing the derivative of the macroscopic dielectric tensor with respect to the normal mode coordinate[12].

## Supplementary Tables

**Table S1.** Lattice parameters of *I4/mcm* MgO$_2$

| Run # | P, GPa | a, Å | σ (a, Å) | c, Å | σ (c, Å) | V, (Å3) | σ (V, Å3) |
|---|---|---|---|---|---|---|---|
| A1 | 93.5 | 3.9994 | 0.0009 | 4.7458 | 0.0032 | 75.9104 | 0.0482 |
| A1 | 94.0 | 3.9988 | 0.0010 | 4.7343 | 0.0031 | 75.7028 | 0.0417 |
| A1 | 95.5 | 3.9973 | 0.0014 | 4.7271 | 0.0034 | 75.5300 | 0.0501 |
| A1 | 95.7 | 3.9961 | 0.0008 | 4.7243 | 0.0038 | 75.4409 | 0.0501 |
| A2 | 101.5 | 3.9787 | 0.0001 | 4.7188 | 0.0004 | 74.6971 | 0.0058 |
| A2 | 102.5 | 3.9803 | 0.0001 | 4.7115 | 0.0004 | 74.6421 | 0.0058 |
| B2 | 130.0 | 3.9211 | 0.0007 | 4.6040 | 0.0027 | 70.7882 | 0.0375 |
| B2 | 139.0 | 3.8985 | 0.0007 | 4.5605 | 0.0027 | 69.3124 | 0.0363 |
| B2 | 154.0 | 3.8666 | 0.0013 | 4.5646 | 0.0038 | 68.2455 | 0.0559 |
| B2* | 152.0 | 3.8862 | 0.0009 | 4.5502 | 0.0028 | 68.7188 | 0.0362 |
| B2* | 150.0 | 3.8909 | 0.0014 | 4.5849 | 0.0033 | 69.4122 | 0.0458 |
| B2* | 146.5 | 3.9060 | 0.0008 | 4.5529 | 0.0035 | 69.4645 | 0.0535 |
| B2* | 116.5 | 3.9451 | 0.0010 | 4.6359 | 0.0030 | 72.1509 | 0.0450 |
| B2* | 115.0 | 3.9666 | 0.0009 | 4.6063 | 0.0028 | 72.4751 | 0.0385 |
| B2* | 114.5 | 3.9619 | 0.0010 | 4.6631 | 0.0031 | 73.1970 | 0.0459 |
| B2* | 112.0 | 3.9617 | 0.0013 | 4.6617 | 0.0039 | 73.1655 | 0.0612 |
| B2* | 110.0 | 3.9800 | 0.0010 | 4.6614 | 0.0031 | 73.8394 | 0.0463 |
| B2* | 108.0 | 3.9718 | 0.0013 | 4.6822 | 0.0040 | 73.8614 | 0.0620 |
| B2* | 103.0 | 3.9831 | 0.0014 | 4.7070 | 0.0040 | 74.6762 | 0.0630 |
| B2* | 83.5 | 4.0478 | 0.0012 | 4.8181 | 0.0034 | 78.9424 | 0.0520 |
| B2* | 83.0 | 4.0299 | 0.0014 | 4.8199 | 0.0042 | 78.2748 | 0.0676 |
| B2* | 82.5 | 4.0517 | 0.0012 | 4.8096 | 0.0034 | 78.9547 | 0.0519 |
| B2* | 74.0 | 4.0784 | 0.0019 | 4.8687 | 0.0040 | 80.9809 | 0.0585 |

Decompression run is marked with an asterisk

**Table S2.** Computed lattice parameters of *I4/mcm* MgO$_2$

| P, GPa | a, Å | c, Å |
|---|---|---|
| 70 | 4.0853 | 4.8762 |
| 80 | 4.0572 | 4.8277 |
| 90 | 4.0315 | 4.7836 |
| 100 | 4.0076 | 4.7409 |
| 110 | 3.9855 | 4.7021 |
| 120 | 3.9649 | 4.6662 |
| 130 | 3.9455 | 4.6325 |
| 140 | 3.9272 | 4.6004 |
| 150 | 3.9101 | 4.5697 |

**Table S3.** Parameters of the 300 K third-order Birch-Murnaghan EOS of *I4/mcm* MgO$_2$

|  | $V_0$, Å$^3$ | $K_0$, GPa | $K_0`$ |
|---|---|---|---|
| Compression | 108.8 | 127.2 | 4 |
| Decompression | 113 | 110.4 | 4 |
| DFT | 106.8 | 147 | 4.1 |
| MgO (Ref.[13]) | 74.71 | 160.2 | 3.99 |

**Supplementary Figures**

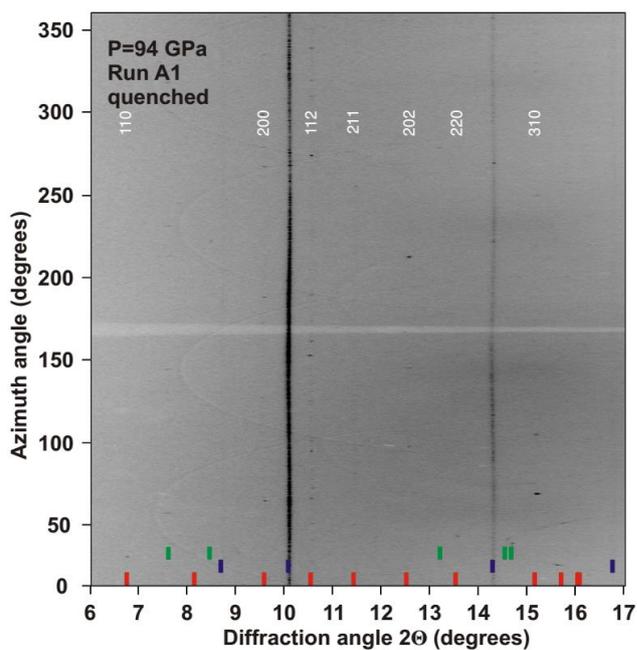

*Figure S1.* XRD image (cake) of I4/mcm $MgO_2$ synthesized at 94 GPa from the mixture of MgO and $O_2$. Red and violet ticks correspond to the positions of I4/mcm $MgO_2$ and MgO, respectively. Green ticks represent spotty reflections of $\zeta$-$O_2$. White labels are Miller indices of the indexed tetragonal phase. A slight curvature of the vertical lines (originating from MgO is due to a pressure gradient in the sample cavity). The x-ray wavelength is 0.3344 Å.

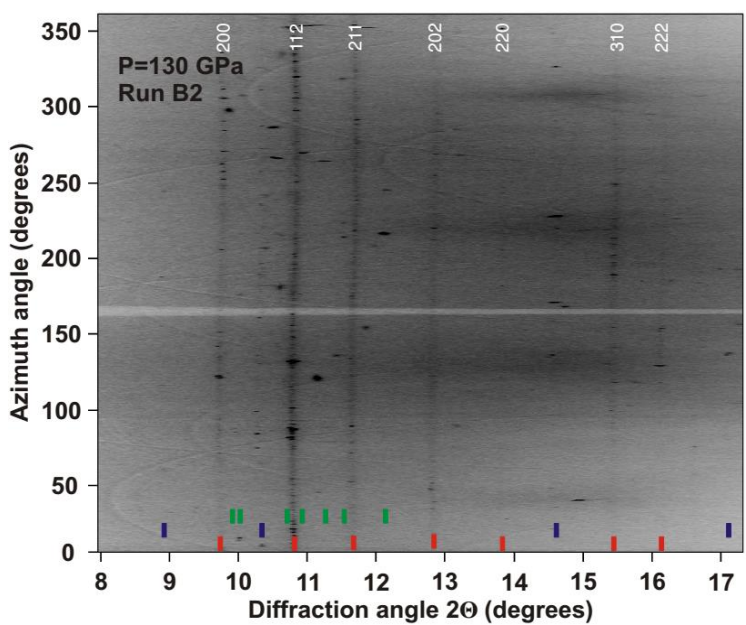

*Figure S2.* XRD image of I4/mcm $MgO_2$ powder synthesized at 130 GPa (seen as dark grey vertical lines) in rectangular coordinates (cake). Red and violet ticks correspond to the positions of I4/mcm $MgO_2$ and MgO, respectively. Green ticks represent some reflections of $\zeta$-$O_2$ (high angle Bragg reflections are not shown). White labels are Miller indices of the indexed tetragonal phase. Part of this XRD pattern ($2\theta$=9-13.5) was used to Rietveld refine the predicted structure of I4/mcm $MgO_2$. The x-ray wavelength is 0.3344 Å.

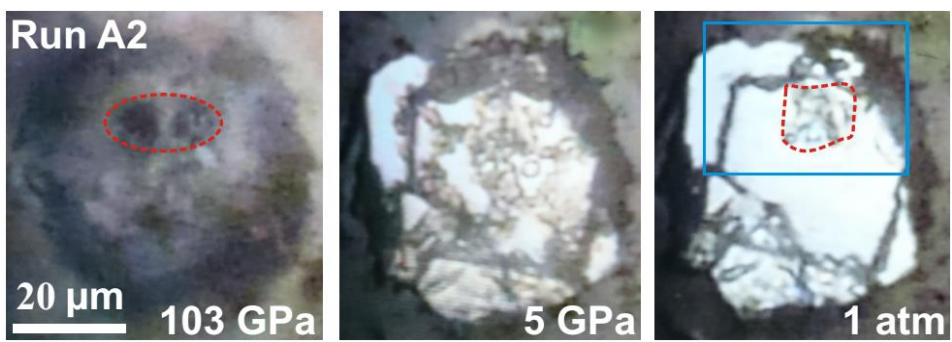

*Figure S3.* Optical images of the A2 sample in a DAC cavity. Red dashed line marks the laser-heated region. Blue solid line corresponds to the sample area shown in the Supplementary Figure S4.

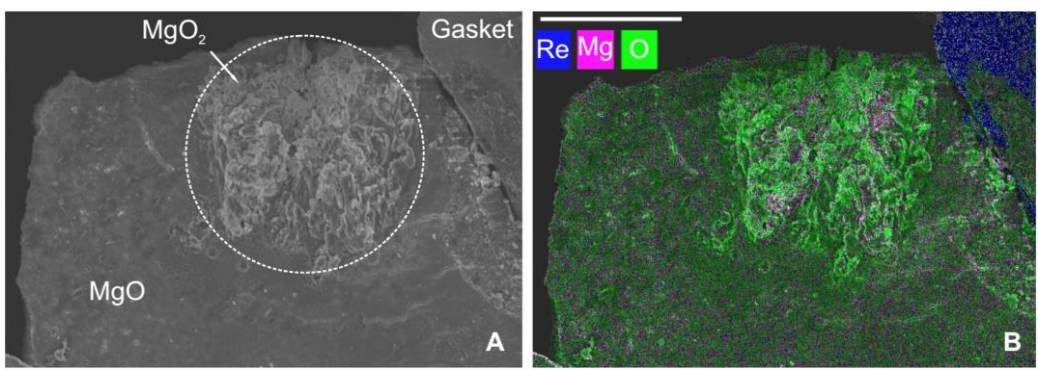

*Figure S4.* Electron microscope images of the extracted sample (run A2). (**A**) SEM micrograph. Laser-heated area is shown with a dashed circle. (**B**) Energy-dispersive x-ray spectroscopy image. Color intensity is proportional to the element abundance. The laser-heated area (white dashed line) has higher oxygen content.

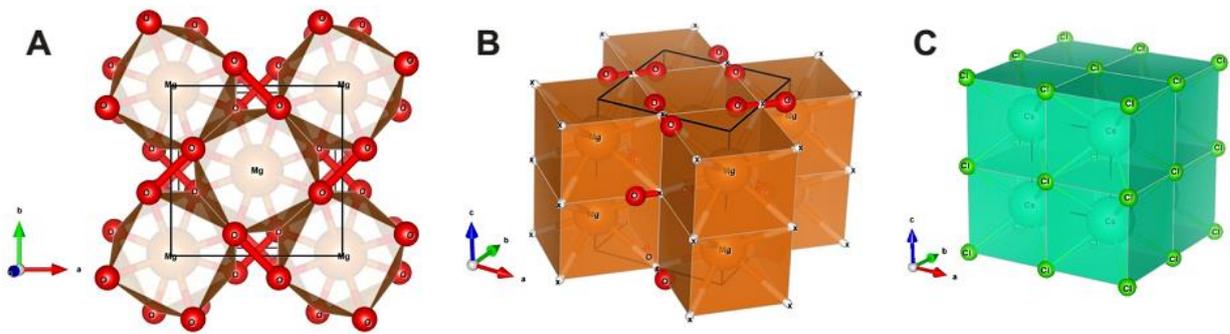

*Figure S5. The crystal structure of I4/mcm MgO$_2$ (panels **A** and **B**) as compared to CsCl-type (**C**). Lattice positions marked with an x in panel B are centers of the O-O dumbbells. Black solid lines represent unit cells.*

**Supplementary References**